\documentclass[conference]{IEEEtran}
\IEEEoverridecommandlockouts

\usepackage{amsmath,amssymb}
\usepackage{booktabs}
\usepackage{graphicx}
\usepackage{cite}
\usepackage[hidelinks]{hyperref}
\usepackage{acronym}
\usepackage{xcolor}
\usepackage{tikz}

\acrodef{10G-EPON}{10 Gigabit EPON}
\acrodef{3GPP}{3rd Generation Partnership Project}
\acrodef{5G}{Fifth Generation}
\acrodef{6G}{Sixth Generation}
\acrodef{A-CPI}{A Controller Plane Interface}
\acrodef{A-RoF}{Analog Radio-over-Fiber}
\acrodef{ABNO}{Application-Based Network Operations}
\acrodef{ADC}{Analog-to-Digital Converter}
\acrodef{AE}{Allocative Efficiency}
\acrodef{amc}{adaptive modulation and coding}
\acrodef{AON}{Active Optical Network}
\acrodef{AI}{Artificial Intelligence}
\acrodef{ML}{Machine Learning}
\acrodef{API}{Application Programming Interface}
\acrodef{AWG}{Arrayed Waveguide Grating}
\acrodef{B2B}{Business to Business}
\acrodef{B2C}{Business to Consumer}
\acrodef{B-RAS}{Broadband Remote Access Servers}
\acrodef{BaaS}{Blockchain as a Service}
\acrodef{BB}{Budget Balance}
\acrodef{BBF}{BroadBand Forum}
\acrodef{BBU}{Base Band Unit}
\acrodef{BER}{Bit Error Rate}
\acrodef{BEREC}{Body of European Regulators for Electronic Communications}
\acrodef{BF}{beamforming}
\acrodef{BGP}{Border Gateway Protocol}
\acrodef{BMap}{Bandwidth Map}
\acrodef{BPM}{Business Process Management}
\acrodef{BS}{Base Station}
\acrodef{BSC}{Base Station Controller}
\acrodef{BtB}{Back to Back}
\acrodef{BW}{Bandwidth}
\acrodef{C2C}{Consumer to Consumer}
\acrodef{C-RoFN}{Cloud-based Radio over optical Fiber Network}
\acrodef{CA}{Certificate Authority}
\acrodef{CapEx}{Capital Expenditure}
\acrodef{CDMA}{code division multiple access}
\acrodef{CDR}{Call Drop Rate}
\acrodef{CFO}{carrier frequency offset}
\acrodef{CIR}{Committed Information Rate}
\acrodef{CO}{Central Office}
\acrodef{CoMP}{Coordinated Multipoint}
\acrodef{COP}{Control Orchestration Protocol}
\acrodef{CORD}{Central Office Rearchitected as a Data Centre}
\acrodef{CPE}{Customer Premises Equipment}
\acrodef{CPRI}{Common Public Radio Interface}
\acrodef{CPU}{Central Processing Unit}
\acrodef{CQI}{channel quality indicator}
\acrodef{CR}{cognitive radio}
\acrodef{CSI}{channel state information}
\acrodef{D2D}{Device to Device}
\acrodef{D-CPI}{D Controller Plane Interface}
\acrodef{D-RoF}{Digital Radio-over-Fiber}
\acrodef{DAC}{Digital-to-Analog Converter}
\acrodef{DAS}{Distributed Antenna Systems}
\acrodef{DBA}{Dynamic Bandwidth Allocation}
\acrodef{DBRu}{Dynamic Bandwidth Report upstream}
\acrodef{DC}{Direct Current}
\acrodef{DLT}{Distributed Ledger Technology}
\acrodef{DMT}{Discrete Multitone}
\acrodef{DNS}{Domain Name Server}
\acrodef{DPDK}{Data Plane Development Kit}
\acrodef{DSB}{Double Side Band}
\acrodef{DSL}{Digital Subscriber Line}
\acrodef{DSLAM}{Digital Subscriber Line Access Multiplexer}
\acrodef{DSP}{Digital Signal Processing}
\acrodef{DSS}{Distributed Synchronization Service}
\acrodef{DU}{Distributed Unit}
\acrodef{DUE}{D2D user equipment}
\acrodef{DWDM}{Dense Wave Division Multiplexing}
\acrodef{E-CORD}{Enterprise CORD}
\acrodef{EAL}{Environment Abstraction Layer}
\acrodef{EDF}{Erbium-Doped Fiber}
\acrodef{EFM}{Ethernet in the First Mile}
\acrodef{EMBS}{Elastic Mobile Broadband Service}
\acrodef{EON}{Elastic Optical Network}
\acrodef{EPC}{Evolved Packet Core}
\acrodef{EPON}{Ethernet Passive Optical Network}
\acrodef{eMBB}{enhanced Mobile Broadband}
\acrodef{ETSI}{European Telecommunications Standards Institute}
\acrodef{EVM}{Error Vector Magnitude}
\acrodef{FANS}{Fixed Access Network Sharing}
\acrodef{FASA}{Flexible Access System Architecture}
\acrodef{FBMC}{Filterbank Multicarrier}
\acrodef{FCC}{Federal Communications Commission}
\acrodef{FDD}{Frequency Division Duplex}
\acrodef{FDMA}{frequency-division multiple access}
\acrodef{FEC}{Forward Error Correction}
\acrodef{FFR}{Fractional Frequency Reuse}
\acrodef{FFT}{Fast Fourier Transform}
\acrodef{FPGA}{Field-Programmable Gate Array}
\acrodef{FSO}{Free-Space-Optics}
\acrodef{FSR}{Free Spectral Range}
\acrodef{FTN}{faster-than-nyquist}
\acrodef{FTTC}{Fiber-to-the-Curb}
\acrodef{FTTH}{Fiber-to-the-Home}
\acrodef{FTTx}{Fiber-to-the-x}
\acrodef{G.Fast}{Fast Access to Subscriber Terminals}
\acrodef{GEM}{G-PON encapsulation method}
\acrodef{GFDM}{Generalized Frequency Division Multiplexing}
\acrodef{GMPLS}{Generalized Multi-Protocol Label Switching}
\acrodef{GPON}{Gigabit Passive Optical Network}
\acrodef{GPP}{General Purpose Processor}
\acrodef{GPRS}{General Packet Radio Service}
\acrodef{GSM}{Global System for Mobile Communications}
\acrodef{GTP}{GPRS Tunneling Protocol}
\acrodef{GUI}{Graphical User Interface}
\acrodef{HA}{Hardware Accelerator}
\acrodef{HARQ}{Hybrid-Automatic Repeat Request}
\acrodef{HD}{half duplex}
\acrodef{HetNet}{heterogeneous networks}
\acrodef{IaaS}{Infrastructure as a Service}
\acrodef{I-CPI}{I Controller Plane Interface}
\acrodef{I2RS}{Interface 2 Routing System}
\acrodef{IA}{Interference Alignment}
\acrodef{IAM}{Identity and Access Management}
\acrodef{IBFD}{in-band full duplex}
\acrodef{IC}{Incentive Compatibility}
\acrodef{ICIC}{inter-cell interference coordination}
\acrodef{ICT}{information and communications technology}
\acrodef{IEEE}{Institute of Electrical and Electronics Engineers}
\acrodef{IETF}{Internet Engineering Task Force}
\acrodef{IF}{Intermediate Frequency}
\acrodef{IFFT}{inverse FFT}
\acrodef{ITS}{Intelligent Transport Systems}
\acrodef{InP}{Infrastructure Provider}
\acrodef{IoT}{Internet of Things}
\acrodef{IP}{Internet Protocol}
\acrodef{IQ}{In-phase/Quadrature}
\acrodef{IR}{Individual Rationality}
\acrodef{IRC}{Interference Rejection Combining}
\acrodef{ISI}{inter-symbol interference}
\acrodef{ISP}{Internet Service Provider}
\acrodef{BFT}{Byzantine-Fault-Tolerant}
\acrodef{IV}{Intelligent Vehicle}
\acrodef{ITU}{International Telecommunication Union}
\acrodef{KVM}{Kernel-based Virtual Machine}
\acrodef{L2}{Layer-2}
\acrodef{L3}{Layer-3}
\acrodef{LAN}{Local Area Network}
\acrodef{LO}{Local Oscillator}
\acrodef{LOS}{line-of-sight}
\acrodef{LR-PON}{Long Reach PON}
\acrodef{LSP-DB}{Label Switched Path Database}
\acrodef{LTE-A}{Long Term Evolution Advanced}
\acrodef{LTE}{Long Term Evolution}
\acrodef{M-CORD}{Mobile CORD}
\acrodef{M-MIMO}{massive MIMO}
\acrodef{M2M}{machine-to-machine}
\acrodef{MAC}{Medium Access Control}
\acrodef{ME}{Merging Engine}
\acrodef{MIMO}{multiple input multiple output}
\acrodef{MME}{Mobility Management Entity}
\acrodef{mmWave}{millimeter wave}
\acrodef{MPLS}{Multiprotocol Label Switching}
\acrodef{MRC}{Maximum Ratio Combining}
\acrodef{MSC}{Mobile Switching Centre}
\acrodef{MSP}{Membership Service Providers}
\acrodef{MSR}{Multi-Stratum Resources}
\acrodef{MTC}{machine type communication}
\acrodef{mMTC}{massive Machine Type Communications}
\acrodef{MVNO}{Mobile Virtual Network Operator}
\acrodef{MVNP}{mobile virtual network provider}
\acrodef{MZI}{Mach-Zehnder Interferometer}
\acrodef{MZM}{Mach-Zehnder Modulator}
\acrodef{NE}{Network Element}
\acrodef{NFV}{Network Function Virtualization}
\acrodef{NFVaaS}{Network Function Virtualization as a Service}
\acrodef{NG-PON2}{Next-Generation Passive Optical Network 2}
\acrodef{NGA}{Next Generation Access}
\acrodef{NLOS}{non-line-of-sight}
\acrodef{NMS}{Network Management System}
\acrodef{NRZ}{Non Return-to-Zero}
\acrodef{NTT}{Nippon Telegraph and Telephone}
\acrodef{OBPF}{Optical BandPass Filter}
\acrodef{OBSAI}{Open Base Station Architecture Initiative}
\acrodef{ODN}{Optical Distribution Network}
\acrodef{OFC}{Optical Frequency Comb}
\acrodef{OFDM}{orthogonal frequency-division multiplexing}
\acrodef{OFDMA}{orthogonal frequency-division multiple access}
\acrodef{OLO}{Other Licensed Operator}
\acrodef{OLT}{Optical Line Terminal}
\acrodef{ONF}{Open Networking Foundation}
\acrodef{ONU}{Optical Network Unit}
\acrodef{OOB}{out-of-band}
\acrodef{OOK}{On-off Keying}
\acrodef{OpenCord}{Central Office Re-Architected as a Data Center}
\acrodef{OpEx}{Operating Expenditure}
\acrodef{OSI}{Open Systems Interconnection}
\acrodef{OSS}{Operations Support Systems}
\acrodef{OTT}{Over-the-Top}
\acrodef{OXM}{OpenFlow Extensible Match}
\acrodef{P2MP}{Ethernet over point-to-multipoint }
\acrodef{P2P}{Point-to-Point }
\acrodef{PAM}{Pulse Amplitude Modulation}
\acrodef{PAPR}{Peak-to-Average Power Ratio}
\acrodef{PBMA}{Priority Based Merging Algorithm}
\acrodef{PCE}{Path Computation Elements}
\acrodef{PCEP}{Path Computation Element Protocol}
\acrodef{PD}{Photodiode}
\acrodef{PDCP}{RD Control Protocol}
\acrodef{PDF}{probability distribution function}
\acrodef{PGW}{Packet Gateway}
\acrodef{PHY}{physical Layer}
\acrodef{PIR}{Peak Information Rate}
\acrodef{PMD}{Polarization Division Multiplexing}
\acrodef{PON}{Passive Optical Network}
\acrodef{POTS}{Plain Old Telephone Service}
\acrodef{PoET}{Proof of Elapsed Time}
\acrodef{PoW}{Proof of Work}
\acrodef{PoS}{Proof of Stake}
\acrodef{PTP}{Precision Time Protocol}
\acrodef{PWM}{Pulse Width Modulation}
\acrodef{QAM}{Quadrature Amplitude Modulation}
\acrodef{QoE}{Quality of Experience}
\acrodef{QoS}{Quality of Service}
\acrodef{QPSK}{Quadrature Phase Shift Keying}
\acrodef{R-CORD}{Residential CORD}
\acrodef{RA}{Resource Allocation}
\acrodef{RAN}{Radio Access Network}
\acrodef{RAT}{Radio Access Technology}
\acrodef{RB}{resource block}
\acrodef{RFIC}{radio frequency integrated circuit}
\acrodef{RLC}{Radio Link Control}
\acrodef{RN}{Remote Node}
\acrodef{ROADM}{Reconfigurable Optical Add Drop Multiplexer}
\acrodef{RoF}{Radio-over-Fiber}
\acrodef{RRC}{Radio Resource Control}
\acrodef{RRH}{Remote Radio Head}
\acrodef{RRPH}{Remote Radio and PHY Head}
\acrodef{RRU}{Remote Radio Unit}
\acrodef{RSOA}{Reflective Semiconductor Optical Amplifier}
\acrodef{RU}{radio unit}
\acrodef{Rx}{receiver}
\acrodef{SC-FDMA}{single carrier frequency division multiple access}
\acrodef{SCM}{single carrier modulation}
\acrodef{SD-RAN}{Software Defined Radio Access Network}
\acrodef{SDAN}{Software Defined Access Network}
\acrodef{SDMA}{Semi-Distributed Mobility Anchoring}
\acrodef{SDN}{Software Defined Networking}
\acrodef{SDR}{Software Defined Radio}
\acrodef{SFBD}{Single Fiber Bi-Direction}
\acrodef{SGW}{Serving Gateway}
\acrodef{SI}{self-interference}
\acrodef{SIC}{self-interference cancellation}
\acrodef{SIMO}{Single Input Multiple Output}
\acrodef{SLA}{Service Level Agreement}
\acrodef{SMF}{Single Mode Fiber}
\acrodef{SNMP}{Simple Network Management Protocol}
\acrodef{SNR}{Signal-to-Noise Ratio}
\acrodef{SP}{Service Providers}
\acrodef{Split-PHY}{Split Physical Layer}
\acrodef{SRI-OV}{Single Root Input/Output Virtualization}
\acrodef{SU}{secondary user}
\acrodef{SUT}{System Under Test}
\acrodef{T-CONT}{Transmission Container}
\acrodef{TC}{Transmission Convergence}
\acrodef{TCO}{Total Cost of Ownership}
\acrodef{TD-LTE}{Time Division LTE}
\acrodef{TDD}{Time Division Duplex}
\acrodef{TDM}{Time Division Multiplexing}
\acrodef{TDMA}{Time Division Multiple Access}
\acrodef{TED}{Traffic Engineering Database}
\acrodef{TEID}{Tunnel endpoint identifier}
\acrodef{TLS}{Transport Layer Security}
\acrodef{TPS}{Transactions per Second}
\acrodef{TTI}{transmission time interval}
\acrodef{TWDM}{Time and Wavelength Division Multiplexing}
\acrodef{Tx}{transmitter}
\acrodef{UD-CRAN}{Ultra-Dense Cloud Radio Access Network}
\acrodef{UDP}{User Datagram Protocol}
\acrodef{UE}{user equipment}
\acrodef{UF-OFDM}{Universally Filtered OFDM}
\acrodef{UFMC}{Universally Filtered Multicarrier}
\acrodef{UL}{Uplink}
\acrodef{UMTS}{Universal Mobile Telecommunications System}
\acrodef{URLLC}{Ultra-Reliable Low-Latency Communication}
\acrodef{USRP}{Universal Software Radio Peripheral}
\acrodef{V2I}{Vehicle to Infrastructure}
\acrodef{V2V}{Vehicle to Vehicle}
\acrodef{vBBU}{virtualized BBU}
\acrodef{vBMap}{Virtual Bandwidth Map}
\acrodef{vBS}{virtual Base station}
\acrodef{VCG}{Vickery-Clarke-Groves}
\acrodef{vCPE}{virtual CPE}
\acrodef{vCPU}{Virtual Central Processing Unit}
\acrodef{vDBA}{virtual DBA}
\acrodef{VDSL2}{Very-high-bit-rate digital subscriber line 2}
\acrodef{VLAN}{Virtual Local Area Network}
\acrodef{VM}{Virtual Machine}
\acrodef{VNF}{Virtual Network Function}
\acrodef{VNI}{visual networking index}
\acrodef{VNO}{Virtual Network Operator}
\acrodef{VNTM}{Virtual Network Topology Manager}
\acrodef{vOLT}{virtual OLT}
\acrodef{VPE}{virtual Provider Edge}
\acrodef{VPN}{Virtual Private Network}
\acrodef{V2X}{Vehicle to Everything}
\acrodef{VTN}{Virtual Tenant Network}
\acrodef{VULA}{Virtual Unblunded Local Access}
\acrodef{WAN}{Wide Area Network}
\acrodef{WAP}{Wireless Access Point}
\acrodef{WCDMA}{wide-band code division multiple access}
\acrodef{WDM-PON}{Wavelength Division Multiplexing - Passive Optical Network}
\acrodef{WDM}{Wavelength Division Multiplexing}
\acrodef{WiMAX}{Worldwide Interoperability for Microwave Access}
\acrodef{WRPR}{Wired-to-RF Power Ratio}
\acrodef{XG-PON}{10 Gigabit PON}
\acrodef{XGEM}{XG-PON encapsulation method}
\acrodef{XOS}{XaaS Operating System}
\acrodef{RIC}{RAN Intelligent Controller}
\acrodef{Near-RT RIC}{near-real-time RAN Intelligent Controller}
\acrodef{Near-RT-RIC}{near-real-time RAN Intelligent Controller}
\acrodef{Non-RT RIC}{Non Real Time RAN Intelligent Controller}
\acrodef{Non-RT-RIC}{Non Real Time RAN Intelligent Controller}
\acrodef{xApp}[xApp]{RAN control application}
\acrodef{rApp}{Remote Application}
\acrodef{dApp}{Distributed Application}
\acrodef{CMS}{Conflict Management System}
\acrodef{EG}{Eisenberg--Gale}
\acrodef{NSWF}{Nash Social Welfare Function}
\acrodef{CMC}{conflict mitigation component}
\acrodef{CDC}{conflict detection component}
\acrodef{PMon}{performance monitor}
\acrodef{RCP}{recently changed parameters}
\acrodef{PGD}{Parameter Group Definition}
\acrodef{RCPG}{Recently Changed Parameter Group}
\acrodef{DCKD}{Decision Correlated with KPI Degradation} 
\acrodef{KDO}{KPI Degradation Occurrences}
\acrodef{MNO}{Mobile Network Operator}
\acrodef{SDL}{Shared Data Layer}
\acrodef{SON}{Self Organising Network}
\acrodef{SF}{SON Function}
\acrodef{AM}{Arithmetic Mean}
\acrodef{PKR}{Parameter and KPI Ranges}
\acrodef{KPI}{key performance indicator}
\acrodef{ICP}{input control parameter}
\acrodef{RF}{Radio Frequency}
\acrodef{C-RAN}{Cloud Radio Access Network}
\acrodef{RL}{Reinforcement Learning}
\acrodef{MARL}{Multi Agent Reinforcement Learning}
\acrodef{CCO}{coverage and capacity optimization}
\acrodef{ES}{energy saving}
\acrodef{MRO}{mobility robustness optimization}
\acrodef{MLB}{mobility load balancing}
\acrodef{V-RAN}{Virtualized Radio Access Network}
\acrodef{O-RAN}{Open Radio Access Network}
\acrodef{D-RAN}{Distributed Radio Access Network}
\acrodef{5GB}{5G and Beyond}
\acrodef{O-RU}{Open Radio Unit}
\acrodef{O-DU}{Open Distribution Unit}
\acrodef{O-CU}{Open Centralised Unit}
\acrodef{O-CU-CP}{Open Centralised Unit - Control Plane}
\acrodef{O-CU-UP}{Open Centralised Unit - User Plane}
\acrodef{D-SON}{Distributed SON}
\acrodef{C-SON}{Centralised SON}
\acrodef{CIO}{Cell Individual Offset}
\acrodef{CBR}{Call Block Ratio}
\acrodef{HOR}{Handover Ratio}
\acrodef{CAN}{Cognitive Autonomous Network}
\acrodef{CF}{Cognitive Function}
\acrodef{QACM}{QoS-aware xApp conflict mitigation}
\acrodef{ANN}{Artificial Neural Network}
\acrodef{PR}{Polynomial Regression}
\acrodef{MSE}{Mean Squared Error}
\acrodef{EVS}{Explained Variance Score}
\acrodef{CS}{Conflict Supervision}
\acrodef{TXP}{Tx Power}
\acrodef{TTT}{Time-To-Trigger}
\acrodef{NL}{Neighbour List}
\acrodef{RET}{remote electrical tilt}
\acrodef{HYS}{Handover Hysteresis}
\acrodef{QACMP}{QACM Priority}
\acrodef{DRL}{Deep Reinforcement Learning}
\acrodef{gNB}{next-generation NodeB}
\acrodef{RSRP}{reference signal received power}
\acrodef{P2K}{parameter-to-KPI}
\acrodef{P2X}{parameter-to-xApp}
\acrodef{K2X}{KPI-to-xApp}
\acrodef{VK}{violated-KPI}
\acrodef{UP}{Uplink}
\acrodef{DL}{Downlink}
\acrodef{AIMLFW}{Open RAN AI/ML framework}
\acrodef{KPM}{Key Performance Measurement}
\acrodef{E2SM}{E2 Service Model}
\acrodef{EI}{Enrichment Information}
\acrodef{NWDAF}{Network Data Analytics Function}
\acrodef{PCF}{Policy Control Function}
\acrodef{PPO}{Proximal Policy Optimization}
\acrodef{GPU}{Graphics Processing Unit}
\acrodef{SIMD}{Single Instruction, Multiple Data}
\acrodef{GNN}{Graph Neural Network}
\acrodef{TL}{Traffic Load}
\acrodef{Bi-LSTM}{Bidirectional Long Short-Term Memory}
\acrodef{SMOTE}{Synthetic Minority Over-sampling}
\acrodef{ECDF}{Empirical Cumulative Distribution Function}
\acrodef{OpenRAN}[Open RAN]{Open Radio Access Network}
\acrodef{CU}{Centralised Unit}
\acrodef{DDQN}{Double Deep Q-Network}
\acrodef{DQN}{Deep Q-Network}
\acrodef{BFP}{Block Floating Point}
\acrodef{FR1}{Frequency Range 1}
\acrodef{CDF}{Cumulative Distribution Function}
\acrodef{KS}{Kolmogorov--Smirnov}
\acrodef{INT}{integrated distance}
\acrodef{SINR}{Signal-to-Interference-plus-Noise Ratio}
\acrodef{HOF}{Handover Failure}
\acrodef{CI}{Confidence Interval}
\acrodef{SD}{Standard Deviation}
\acrodef{GT}{game-theoretic}
\acrodef{FC}{fully connected}
\acrodef{COTS}{Commercial Off-The-Shelf}
\acrodef{LSTM}{Long Short-Term Memory}
\acrodef{SONCO}{SON Coordinator}
\acrodef{TNSM}{IEEE Transactions on Network and Service Management}
\acrodef{LLM}{Large Language Model}
\acrodef{SHAP}{SHapley Additive exPlanations}
\acrodef{RT}{Real Time}
\acrodef{RT-RIC}{Real-Time RAN Intelligent Controller}
\acrodef{GenC}{Conflict Generator}

\makeatletter
\renewcommand\section{\@startsection{section}{1}{\z@}%
  {1.2ex plus 1ex minus 0.3ex}%
  {0.5ex plus 0.7ex minus 0ex}%
  {\normalfont\normalsize\centering\scshape}}
\renewcommand\subsection{\@startsection{subsection}{2}{\z@}%
  {1.2ex plus 1ex minus 0.3ex}%
  {0.5ex plus .4ex minus 0ex}%
  {\normalfont\normalsize\itshape}}
\makeatother

\newcommand{\EvalRuns}{20}
\newcommand{\CalibSamples}{180}
\newcommand{\QBound}{3.36}
\newcommand{\EvalCoverage}{94.7}

\newcommand{\PropThroughput}{80.15}
\newcommand{\PropEnergy}{4396.3}
\newcommand{\PropViolation}{58.2}
\newcommand{\PropTolerate}{4.8}
\newcommand{\PropDefer}{25.2}
\newcommand{\PropMitigate}{70.0}

\newcommand{\PropTolerateCount}{19}
\newcommand{\PropDeferCount}{101}
\newcommand{\PropMitigateCount}{280}
\newcommand{\PropChangedCount}{381}
\newcommand{\PriorityInterventionCount}{400}
\newcommand{\UnsafeTolerate}{0}
\newcommand{\InfeasibleMitigations}{30}

\newcommand{\ThroughputRegressionSeeds}{0}
\newcommand{\EnergyIncreaseSeeds}{0}
\newcommand{\ViolationRegressionSeeds}{0}

\newcommand{\ObsThroughput}{68.40 $\pm$ 1.35}
\newcommand{\ObsEnergy}{4512.7 $\pm$ 1.2}
\newcommand{\ObsViolation}{92.80 $\pm$ 1.40}

\newcommand{\PriThroughput}{76.80 $\pm$ 0.15}
\newcommand{\PriEnergy}{4425.6 $\pm$ 1.1}
\newcommand{\PriViolation}{78.50 $\pm$ 1.85}

\newcommand{\PropThroughputCI}{80.15 $\pm$ 0.10}
\newcommand{\PropEnergyCI}{4396.3 $\pm$ 0.6}
\newcommand{\PropViolationCI}{58.20 $\pm$ 1.40}

\newcommand{\PropObsThroughputDelta}{+11.75 $\pm$ 1.42}
\newcommand{\PropObsEnergyDelta}{-116.4 $\pm$ 1.4}
\newcommand{\PropObsViolationDelta}{-34.60 $\pm$ 1.78}

\newcommand{\PropPriThroughputDelta}{+3.35 $\pm$ 0.19}
\newcommand{\PropPriEnergyDelta}{-29.3 $\pm$ 0.9}
\newcommand{\PropPriViolationDelta}{-20.30 $\pm$ 1.62}

\begin{document}

\title{Confidence-Aware xApp Conflict Triage in Open RAN: A Closed-Loop ns3-oran Evaluation}

\title{
    \vspace{-0cm} 
    \begin{tikzpicture}[remember picture, overlay]
        \node[anchor=north, yshift=-0.5cm] at (current page.north) {\fbox{\parbox{\textwidth}{\centering\small {\color{red}This version of the paper has been submitted for possible publication in an IEEE conference. It is subject to IEEE copyright and may be removed upon request.}}}};
    \end{tikzpicture}
    \vspace{0cm} 
    Confidence-Aware xApp Conflict Triage in Open RAN: A Closed-Loop ns3-oran Evaluation

}


\author{
\IEEEauthorblockN{Abdul Wadud\IEEEauthorrefmark{1}\IEEEauthorrefmark{2},~\IEEEmembership{Graduate Student Member, IEEE}\\
Fatemeh Golpayegani\IEEEauthorrefmark{1},~\IEEEmembership{Senior Member, IEEE}, and
Nima Afraz\IEEEauthorrefmark{1},~\IEEEmembership{Senior Member, IEEE}}
\IEEEauthorblockA{\IEEEauthorrefmark{1}School of Computer Science, University College Dublin, Ireland}
\IEEEauthorblockA{\IEEEauthorrefmark{2}Bangladesh Institute of Governance and Management, Dhaka, Bangladesh}}

\maketitle

\begin{abstract}
In the \ac{OpenRAN}, independent \acp{xApp} may request changes to the same radio parameter to optimize their own KPIs, causing conflicts. When conflicts are detected via \ac{KPI} degradation, current conflict management systems usually resolve them immediately, without considering how long the degradation lasts or whether the network might recover naturally, wasting \ac{RU} energy on interventions that may not have been needed. This paper presents a confidence-aware triage layer that assesses the reliability of its predictions by calibrating intervals from past prediction errors. It also decides, for each conflict, whether to tolerate, defer, or mitigate it based on the degraded \ac{KPI} prediction confidence and persistence of the conflict. It is evaluated on a common conflict between two \ac{OpenRAN} \acp{xApp}: \ac{CCO} and \ac{ES}, which issue opposing requests for transmit power and affect \ac{RSRP}. The triage layer is validated in a closed-loop ns3-oran simulation with 5G-LENA. Results show that the method reduces unnecessary interventions by 30\% and maintains \ac{RU} energy efficiency relative to fixed priority while improving it by 2.6\% over the no-mitigation baseline, while maintaining a statistical safety guarantee for \ac{RSRP}.
\end{abstract}

\begin{IEEEkeywords}
Open RAN, xApp conflict, conflict triage, uncertainty, conformal prediction, ns3-oran
\end{IEEEkeywords}

\acresetall

\section{Introduction}
\IEEEPARstart{T}{he} transition from \ac{5G}-Advanced to \ac{6G} mobile communication systems is often described as a shift from \ac{AI} operating alongside the network to \ac{AI} being directly integrated into its management control cycle and orchestration~\cite{chen2026toward}. \ac{OpenRAN} is the leading architecture enabling this transformation through the concept of a programmable \ac{RIC}. \acp{xApp} and \acp{rApp} are \ac{AI}-based network management agents deployed in the \ac{Near-RT RIC} and \ac{Non-RT RIC}, respectively, are some examples of this transformation. This architecture enables statistical or \ac{AI}-based learned models to issue live control decisions to \ac{RAN} components and supports multi-vendor and AI-native control. However, it also introduces the possibility of independent network management applications, i.e., \acp{xApp}, issuing conflicting control decisions. Consequently, conflict detection and mitigation have emerged as critical areas of research in \ac{OpenRAN}~\cite{adamczyk2023conflict,wadud2023conflict,wadud2024xapp,wadud2026ai}. These challenges underscore two primary requirements for \ac{AI}-native \ac{6G} control. First, each new \ac{AI}-based network management component must be energy- and resource-efficient. Second, \ac{AI}-based network management components must be sufficiently reliable and accountable for carrier-grade networks, not just accurate on average.

The \ac{CDC} identifies and classifies conflicts; afterwards, the \ac{CMC} determines the appropriate mitigation strategy. When a conflict is detected, it does not immediately mean that an intervention is required. In some cases, a conflict may not persist long enough to require mitigation and may self-resolve before intervention, or intervention may come too late, after it has already disappeared. In such cases, the premature mitigation intervention can prevent beneficial xApp operations. On the other hand, excessive delay in taking appropriate mitigation action may cause significant performance degradation and a bad user experience. Therefore, the \ac{CMC} requires an intermediate decision layer between immediate action and indefinite postponement. Previous studies have addressed this challenge through various strategies. For example, scheduling xApp execution to preempt conflicts~\cite{cinemre2025xapp}, allowing application pairs only if they satisfy an operator-defined severity threshold (PACIFISTA~\cite{del2024pacifista}), searching for alternative parameter values based on each xApp's \ac{QoS} predictions (QACM~\cite{wadud2024qacm}), or evaluating each possible action in a network digital twin before selection (COMIX~\cite{giannopoulos2025comix}) are some significant works to be highlighted. In our earlier work~\cite{wadud2026cost}, we introduced a cost-aware conflict triage approach that selects whether to tolerate, defer, or mitigate a detected conflict based on its expected mitigation benefit and operational cost. However, that approach does not explicitly evaluate whether a predicted \ac{KPI} satisfies an operator-defined constraint under prediction uncertainty. In this paper, we address this aspect by introducing confidence-aware triage using calibrated prediction intervals to guide decisions. Unlike the earlier simulator-driven evaluation, this work also evaluates the proposed method in a closed-loop ns3-oran environment.

These approaches either prioritise one conflicting \ac{xApp} action over the others or attempt to balance them. They do not answer the immediate question of how the mitigation controller should respond when two live control decisions conflict at a particular timestamp and the next-control-cycle outcome is uncertain. This work addresses this gap by leveraging current control-decision values and the latest cell measurements. It also considers a \ac{MNO}-defined \ac{QoS} threshold and a prediction-error interval obtained from separate independent seed results. However, the final mitigation decision remains reactive, as the \ac{CDC} has already detected the conflict. We use prediction only to inform the mitigation response, not to detect conflicts. 

This paper presents three primary contributions. First, it introduces a confidence-aware conflict triage method as a reactive decision layer between the \ac{CDC} and the \ac{CMC} utilizing the conflict report generated by the \ac{CDC}. Second, it defines a three-way triage model for tolerating, deferring, or mitigating each control-decision conflict batch. Each control decision conflict batch is formed for the same cell, parameter, and control cycle. We employ a straightforward response model and empirically measure its statistical reliability during each control cycle, so the response model remains predictable even when predictions are imperfect. Third, the paper evaluates this triage model in a practical scenario involving conflicts over \ac{TXP} between the \ac{CCO} and \ac{ES} \acp{xApp} within the \ac{Near-RT RIC} control-decision cycle. We evaluate using closed-loop tests that use separate calibration and evaluation seeds. Finally, the network \acp{KPI} is reported, including the \ac{RU} energy savings achieved by the reactive layer, and the circumstances in which the method is less effective.


\section{System Model}
In this section, we describe the key components involved in the conflict mitigation process, which includes  \acp{xApp} that issue control decisions, the cells and control parameters,  the involved \acp{KPI}, and the conflict report passed from the \ac{CDC} to the triage layer of the \ac{CMC}. Let $\mathcal X=\{x_1,\ldots,x_M\}$ be the set of \acp{xApp}, where $x_i\in\mathcal X$ is a single \ac{xApp}. Let $\mathcal N$, $\mathcal P$, and $\mathcal K$ represent the sets of controlled cells, \acp{ICP}, and \acp{KPI}. Let $\mathcal C$ be the set of \ac{xApp} control decisions. Each control decision $c_i\in\mathcal C$ is $c_i=(x_i,n_i,p_i,\Delta_i,t_i)$, where $x_i\in\mathcal X$ is the source \ac{xApp}, $n_i\in\mathcal N$ is the target cell, $p_i\in\mathcal P$ is the \ac{ICP}, $\Delta_i$ is the requested change, and $t_i$ is the control epoch. The control cycle lasts $\Delta t$.

Conflict detection is not a contribution of this paper. The triage method we propose operates strictly after a conflict has already been detected, reusing the conflict management system from our earlier work~\cite{wadud2024xapp,wadud2026ai}. In that system, the \ac{PMon}, \ac{CDC}, and \ac{CMC} are separate modules in the \ac{Near-RT RIC}. The \ac{RIC} database keeps records of \ac{RCP}, \ac{P2X}, \ac{P2K}, \ac{K2X}, and \ac{VK}. \Ac{P2X}, \ac{P2K}, and \ac{K2X} show known xApp dependencies, while \ac{RCP} and \ac{VK} provide timestamped runtime data. At each epoch $t$, the control-decision collector groups requests by target cell and \ac{ICP}: $\mathcal B_{n,p,t}=\{c_i\in\mathcal C:n_i=n,\ p_i=p,\ t_i=t\}$. The \ac{CDC} uses the conflict catalog and runtime records to detect and classify conflicts. There are three types of conflicts: direct conflicts (incompatible requests from different xApps for the same ICP and scope), indirect conflicts (a changed \ac{ICP} linked to a degraded \ac{KPI} owned by another \ac{xApp} through \ac{P2K} and \ac{K2X}), and implicit conflicts (an observed \ac{KPI} effect not found in the known \ac{P2K} relation).

After classification, the CDC creates a conflict report $e_t=(t,n,p,z,\mathcal X_c,\mathcal X_a,\mathcal K_v,\mathcal B_{n,p,t})$. Here, $z$ is the conflict class, $\mathcal X_c\subseteq\mathcal X$ lists the conflicting xApps, $\mathcal X_a\subseteq\mathcal X$ lists affected xApps, and $\mathcal K_v\subseteq\mathcal K$ lists affected or violated KPIs. In this paper, $\mathcal X$ includes four xApps often used in Open RAN: \ac{MRO}, \ac{MLB}, \ac{CCO}, and \ac{ES}. The CDC is available to all four xApps. Our triage evaluation focuses on a common conflict between two of them: \ac{CCO} and \ac{ES}. Both the \acp{xApp} request direct changes to \ac{TXP}, and the proposed triage protects the \ac{RSRP} constraint affected by these changes.

\begin{figure}[t]
\centering
\includegraphics[scale=0.3]{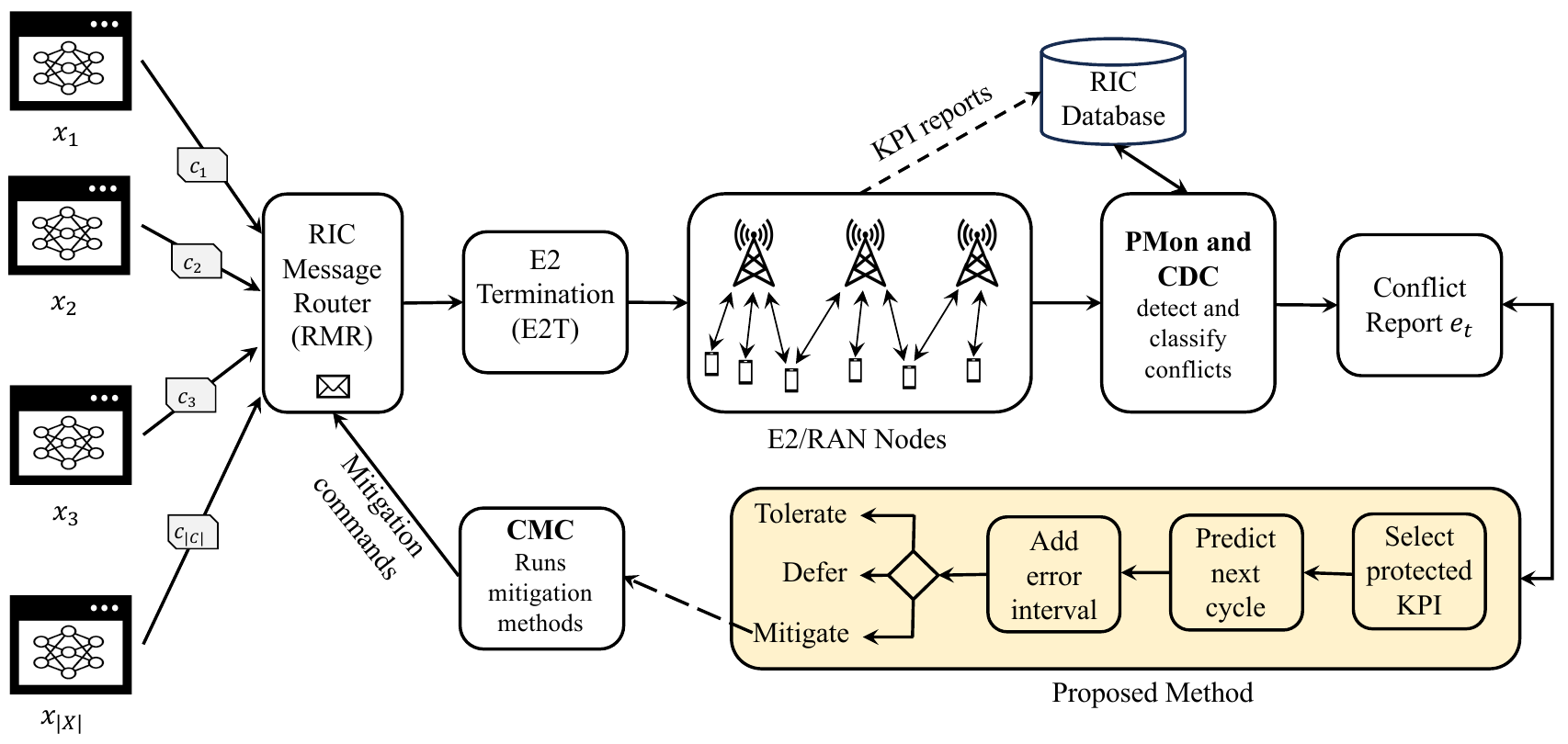}
\caption{Closed-loop conflict-management pipeline. Control decisions go to the E2/RAN nodes through the RIC Message Router (RMR) and E2 termination (E2T). PMon and CDC return $e_t$, and the proposed triage chooses the response before \ac{CMC} runs.}
\label{fig:architecture}
\end{figure}


\section{Confidence-Aware Triage}
\subsection{General framework}
\label{sec:genFrame}
For each conflicting control-decision batch, the triage process predicts the impact on the \acp{KPI} relevant to the operator. It also estimates the potential prediction error for that \ac{KPI} (for instance, \ac{RSRP} in the \ac{CCO}--\ac{ES} case study below; the framework itself is not RSRP-specific). The remainder of this subsection constructs the proposed method into a control decision rule applicable to any KPI and deployment. The proposed framework is compatible and can be deployed with any conflict reports containing $|\mathcal X|$ xApps. The figure~\ref{fig:architecture} shows that the proposed method is a decision layer between the existing \ac{CDC} and \ac{CMC} modules. The proposed method is triggered when the conflict report $e_t$ comes from the \ac{CDC}. 

Let $\mathcal R\subseteq\mathcal P\times\mathcal K$ be a registry of validated \ac{ICP}--\ac{KPI} response models. If a reactive conflict report on control parameter $p$ arrives, the protected set can be defined as $\mathcal K_e=\{k\in\mathcal K_v:(p,k)\in\mathcal R\}$. Here, $\mathcal K_v$ denotes the set of affected or violated \acp{KPI} identified in the conflict report. Similarly, $\mathcal R$ shows that triage has a tested response plan for that \ac{ICP}--\ac{KPI} pair. However, when no confirmed solution is found for a pair, meaning $\mathcal K_e$ is empty, the \ac{CMC} applies an existing mitigation policy. Each $k\in\mathcal K_e$ has a threshold $T_k$ and direction $v_k\in\{1,-1\}$. The value $v_k=1$ refers to a \ac{KPI} that should stay above $T_k$. On the other hand, $v_k=-1$ means one that should stay below it. Given the latest state $\mathbf s_t$ and a candidate control-decision batch $\widetilde{\mathcal B}$, response model $f_{p,k}$ produces the signed service margin
\begin{equation}
 \widehat m_{k,t+1}(\widetilde{\mathcal B})
 =v_k\bigl(f_{p,k}(\mathbf s_t,\widetilde{\mathcal B})-T_k\bigr).
 \label{eq:general_margin}
\end{equation}
A positive margin means the constraint is satisfied, while a negative margin shows it has been violated. The predictor $f_{p,k}$ can be either analytical or learned, but it must use the same input and output format. The prediction margin represents the error as a single best guess. The framework keeps track of how wrong that prediction has been before. For held-out absolute margin residuals $a_{k,j}$ (past predicted errors), the finite-sample conformal radius is
\begin{equation}
 q_{p,k}=a_{k,(h)},\quad
 h=\min\{N_k,\lceil(N_k+1)(1-\alpha_k)\rceil\},
 \label{eq:conformal}
\end{equation}
Here, $a_{k,(h)}$ is the $h$th ordered residual~\cite{lei2018distributionfree}, and $\alpha_k\in(0,1)$ is the error tolerance chosen by the operator (for example, $0.05$ for 95\% coverage). The value $q_{p,k}$ is the largest error found on $(1-\alpha_k)$ of the calibration data. This means that, on average, future errors will not be larger than this radius with probability at least $1-\alpha_k$. The margin interval is $J_{k,t}=[\underline m_{k,t},\overline m_{k,t}]$, where $\underline m_{k,t}=\widehat m_{k,t+1}-q_{p,k}$ and $\overline m_{k,t}=\widehat m_{k,t+1}+q_{p,k}$. This interval shows the predicted value plus or minus the conformal radius. The radius is measured from data, not set by hand.

Let $d_t$ count consecutive deferrals for the same cell and ICP. Each control-decision batch is first classified into one of three states,
\begin{equation}
\begin{aligned}
S_t &\triangleq \bigl(\forall k\in\mathcal K_e:\underline m_{k,t}\ge0\bigr), && \text{(safe)},\\
U_t &\triangleq \bigl(\exists k\in\mathcal K_e:\overline m_{k,t}<0\bigr), && \text{(unsafe)},\\
I_t &\triangleq \neg S_t \land \neg U_t, && \text{(uncertain)},
\end{aligned}
\label{eq:states}
\end{equation}
and the triage decision is
\begin{equation}
A_t=
\begin{cases}
\texttt{tolerate}, & S_t,\\
\texttt{mitigate}, & U_t \lor (I_t \land d_t\ge D_{\max}),\\
\texttt{defer}, & I_t \land d_t<D_{\max}.
\end{cases}
\label{eq:general_triage}
\end{equation}

Equation~\eqref{eq:general_triage} partitions each conflict into three states. A conflict is classified as safe when the pessimistic outcome satisfies all protected KPI constraints, and unsafe when the optimistic outcome violates at least one constraint. Otherwise, the outcome is uncertain. Safe conflicts are tolerated, and unsafe conflicts are mitigated. Uncertain conflicts are deferred while $d_t<D_{\max}$ and are mitigated once the maximum deferral limit is reached. A tolerated batch proceeds unchanged. In contrast, deferral withholds control decisions that would reduce a protected margin and waits for the next KPI report, subject to a maximum of $D_{\max}=\lceil H_{\max}/\Delta t\rceil$ cycles. The estimated statistical error bound relies on the assumption that calibration and future errors are exchangeable. Under this condition, the probability of an unsafe tolerated event for KPI $k$ (where $a_k>q_{p,k}$) is at most $\alpha_k$ across future reports. For multiple protected KPIs, the combined tolerance is bounded by the union $\sum_{k\in\mathcal K_e}\alpha_k$.

If mitigation is required, the CMC finds the smallest valid control-decision change that satisfies every protected KPI constraint:
\begin{equation}
\begin{aligned}
 \widetilde{\mathcal B}^{*}
 &=\arg\min_{\widetilde{\mathcal B}\in\mathcal A(e_t)}
 D_p(\mathcal B_{n,p,t},\widetilde{\mathcal B}),\\
 \mathrm{s.t.}\quad
 &\underline m_{k,t}(\widetilde{\mathcal B})\ge0,
 \quad\forall k\in\mathcal K_e.
\end{aligned}
 \label{eq:general_mitigation}
\end{equation}
In this context, $\mathcal A(e_t)$ denotes the set of valid control-decision changes, and the function $D_p$ quantifies the magnitude of the change in the natural units of the ICP. If no batch satisfies the constraint, the \ac{CMC} implements its predefined safe fallback mitigation action. Such occurrences are recorded. The confidence measure determines whether the available evidence justifies any intervention. The Equation~\eqref{eq:general_mitigation} limits how much intervention can be made. To efficiently utilise the proposed method, the \ac{MNO} must register a protected \ac{KPI}, a validated response model, a set of permissible actions, and a metric for control-decision change. In this paper, we assume that each is set by the \ac{MNO} following \ac{SLA} requirements.


\subsection{CCO and ES TxPower Case Study}
This section discusses a typical conflict scenario to show how the proposed triage model works when implemented. Suppose the  \ac{CCO} increases a gNB’s \ac{TXP} to help boost the signal strength for users with weak signals. In a subsequent request, the \ac{ES} decreases \ac{TXP} to save energy. When both these control decisions are made in the same control cycle, a direct conflict occurs between these two \acp{xApp}. In this section, we consider this direct conflict between \ac{CCO} and \ac{ES} over \ac{TXP} as an example, showing each step from the \ac{CDC} report to the final triage action. The main motivation of this example is to show how the triage benefits the overall conflict mitigation process and avoids unnecessary interventions when \ac{RSRP} is already sufficient.

The \ac{CCO} \ac{xApp} asks for a \ac{TXP} increase only after its low-\ac{RSRP} trigger is reached. Therefore, the \ac{RSRP} is included in $\mathcal K_v$. If \ac{ES} requests to decrease the \ac{TXP} in the same control cycle, the \ac{CDC} reports $e_t$ with $p=\text{TxPower}$, $z=\text{direct}$, $\mathcal X_c=\{\mathrm{CCO},\mathrm{ES}\}$, and both the control decisions. Let us consider that $\Delta^-<0$ is the sum of the negative \ac{TXP} requests, and $\Delta^+>0$ is the sum of the positive requests in this batch. The registry $\mathcal R$ includes the validated \ac{TXP}--\ac{RSRP} response-model pair. We consider that the $\mathcal K_e=\{\mathrm{RSRP}\}$, the \ac{SLA} threshold set by the \ac{MNO} is $T_R$, and $v_R=1$.  Here, the trigger for \ac{CCO} \ac{xApp} and the service threshold $T_R$ have different roles. The former starts the reactive process, while the latter sets the minimum \ac{RSRP} value that triage must protect.

The current state  $\mathbf s_t$ provides the current worst serving-cell \ac{RSRP} called $r_t$. As a first estimate, a 1~dB change in the \ac{TXP} value leads to about a 1~dB change in \ac{RSRP} value in the same direction. The response model then gives $f_{p,R}(\mathbf s_t,\mathcal B)=r_t+\Delta^-+\Delta^+$. Therefore, the future cycle \ac{RSRP} value is the current \ac{RSRP} plus the net \ac{TXP} value change. Applying this to Eq.~\eqref{eq:general_margin} gives the predicted margin $\widehat m_{R,t+1}=r_t+\Delta^-+\Delta^+-T_R$, which shows how much the predicted \ac{RSRP} is above or below the service threshold $T_R$. After that, $q$ is estimated from the held-out errors $|r_{j+1}-\widehat r_{j+1}|$, which are the differences between each past prediction and the actual \ac{RSRP}.  The prediction interval is $I_t=[L_t,U_t]=[\widehat r_{t+1}-q,\widehat r_{t+1}+q]$, and its margin is $J_{R,t}=[L_t-T_R,U_t-T_R]$. Equation~\eqref{eq:general_triage} accepts the batch when $L_t\ge T_R$, mitigates it when $U_t<T_R$, and defers otherwise which is waiting until the delay limit. This is the same decision rule as the general framework presented in Section~\ref{sec:genFrame}.

\begin{figure}[t]
\centering
\includegraphics[scale=0.9]{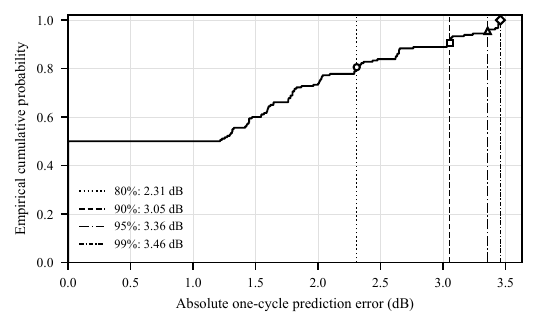}
\vspace{-0.5cm}
\caption{Absolute one-cycle prediction errors on calibration seeds. Markers show the finite-sample radii used in the closed-loop ablation.}
\label{fig:q_calibration}
\end{figure}

When mitigation is needed, the proposed triage method aims to minimise disruptive changes and maximise the overall network benefit. The equation~\eqref{eq:general_mitigation} keeps \ac{CCO}'s control decision unchanged and uses a scale factor $\gamma\in[0,1]$ for \ac{ES}'s control decision. Here, if $\gamma=1$, the full power cut is applied. If $\gamma=0$, the cut is blocked. The largest possible $\gamma$ is $\gamma^*=[(r_t+\Delta^+-q-T_R)/(-\Delta^-)]_0^1$, where $[z]_0^1=\min(1,\max(0,z))$ keeps the result scaled between 0 and 1.  If no $\gamma$ meets the \ac{RSRP} service threshold, $\gamma^*=0$ stops any further coverage reduction until \ac{RSRP} is fixed.

\subsection{Numerical Example}
This section discusses the above theory in a numerical manner so that it becomes clear to understand how the proposed triage method works. Suppose \ac{CCO} asks for $+1$~dB \ac{TXP} and \ac{ES} asks for $-1$~dB in the same cycle. Since these requests are equal and opposite, they cancel each other's action, and hence the predicted \ac{RSRP} becomes the same as the current cycle. For this example, we consider that the \ac{RSRP} threshold is $T_R=-95$~dBm, which leads to the error radius of $q=3.36$~dB for this current conflicting decision pair (see Fig.~\ref{fig:q_calibration}). Applying this value in Eq.~\eqref{eq:general_triage}, we get a mitigation action if $r_t$ is below $-98.36$~dBm. We get a defer and wait if $r_t$ is between $-98.36$ and $-91.64$~dBm because of the uncertainty. We get a tolerate if the $r_t$ is at or above $-91.64$~dBm ($T_R+q$) because even worse prediction error would ensure that the \ac{KPI} meets its target. An important thing to note here that any deferred control decision action that is still unresolved after $D_{\max}$ cycles is eventually handled as a mitigation. Table~\ref{tab:worked_example} shows this rule for three measured RSRP values.

\begin{table}[t]
\centering
\caption{Step-by-step decision for CCO $+1$ dB and ES $-1$ dB.}
\label{tab:worked_example}
\setlength{\tabcolsep}{2.8pt}
\begin{tabular}{cccc}
\toprule
$r_t$ (dBm) & $I_t$ (dBm) & Decision & \ac{TXP} sent\\
\midrule
$-90$ & $[-93.36,-86.64]$ & Tolerate & $+1,-1$ dB\\
$-95$ & $[-98.36,-91.64]$ & Defer & $+1$ dB\\
$-102$ & $[-105.36,-98.64]$ & Mitigate & $+1$ dB\\
\bottomrule
\end{tabular}
\end{table}


We also consider a fourth scenario when a deferred control decision reaches the maximum allowable delay. This situation remains ambiguous as it doesn’t fall into the triage conditions. The proposed method handles this case by proportionally scaling the control decision. For example, if $r_t=-92.5$~dBm and the cell has already reached the maximum number of deferrals ($d_t=D_{\max}$), a decision needs to be made. Using Eq.~\eqref{eq:general_mitigation}, we get $\gamma^*=(-92.5+1-3.36+95)/1=0.14$. So, only 14\% of ES's requested 1~dB power reduction can be allowed. This approach, even in the worst case, helps to save some energy when predicted \ac{RSRP} stays at or above the threshold. The predictor and the value for $q$ estimated here are for this specific scenario used for explaining the proposed method. Each new entry for conflict among other \acp{xApp} in $\mathcal R$ needs its own response model and calibration data.

\begin{figure}[t]
\centering
\includegraphics[width=\columnwidth]{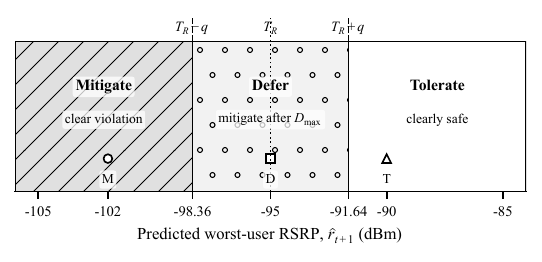}
\vspace{-0.8cm}
\caption{CCO--ES triage regions for $T_R=-95$~dBm and $q=\QBound$~dB. Uncertain control decisions are deferred until the delay limit.}
\label{fig:decision_regions}
\end{figure}


\section{Closed-Loop Evaluation}
We evaluated the proposed framework with a full network simulation using a closed-loop \ac{OpenRAN} environment. Therefore, each policy's decisions directly affect the radio state for later control decisions. The \ac{CDC} and \ac{CMC} paths are implemented in ns3-oran~\cite{ns3oran_github}. The \ac{CDC} creates a conflict record before triage, keeping control-decision associations and avoiding repeated detection, as shown in Figure~\ref{fig:architecture}. The \ac{CMC} filters control decisions before they reach the \ac{gNB}, the 5G base station. The simulation uses ns-3.42 and 5G-LENA v.5~\cite{patriciello2019e2e}, with 13 \acp{gNB} and 117 \ac{UE} devices on a 3.5~GHz, 20~MHz carrier, using numerology 1 (30~kHz subcarrier spacing)~\cite{tgpp38211}. The channel model used in this simulation follows the \ac{3GPP} urban microcell (UMi) specification~\cite{tgpp38901} with either \ac{LOS} or \ac{NLOS} for each link. Each of the four \acp{xApp} deployed in the scenario makes a control decision every $\Delta t=0.5$~s control cycle. Each \ac{UE} gets a steady 20~Mbit/s \ac{UDP} downlink flow, and an EARTH-based \ac{RU} energy model tracks energy use~\cite{auer2011energy, wadud2025ru}. The \ac{SLA} threshold for \ac{RSRP} is set to $T_R=-95$~dBm and can be changed by the operator through the command line. We use 10 different seeds to estimate $q$ and evaluate using 20 other seeds that were not used for estimating  $q$ earlier. We use the same random conditions for each control policy, and all results are reported with a two-sided Student-$t$ 95\% \acp{CI} across seeds.

We compare three mitigation policies. The first is no mitigation, which allows all control decisions irrespective of having conflicts. The second is the priority-based method studied in \cite{adamczyk2023conflict}, which prefers control decisions coming from \ac{xApp} with higher priority. The last one is our proposed method that uses the confidence-aware triage rule from Eq.~\eqref{eq:general_triage}. For the proposed method, we set the $H_{\max}=1$~s, so $D_{\max}=2$ cycles. Each run records the control decisions, conflict records, interval limits, chosen action, $\gamma^*$, RSRP, throughput, energy, and handovers. 

\begin{figure*}[t]
\centering
\includegraphics[width=0.88\textwidth]{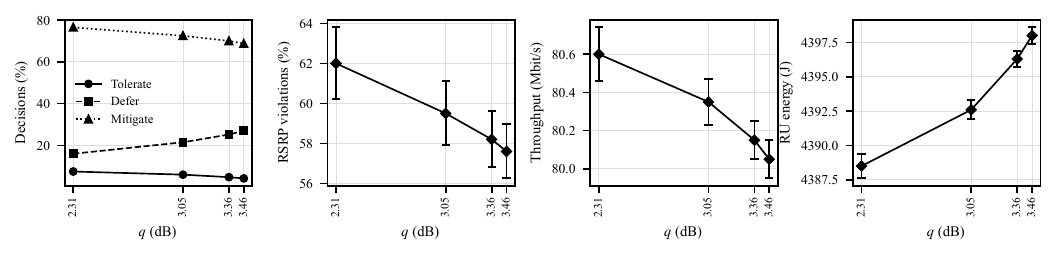}
\vspace{-0.5cm}
\caption{Closed-loop ablation of the conformal radius: (a) actions, (b) decision-time RSRP violations, (c) throughput, and (d) RU energy.}
\vspace{-0.3cm}
\label{fig:q_ablation}
\end{figure*}


\section{Results and Discussion}
This section presents the closed-loop experimental results for the three policies that we are comparing. The evaluation shows how the proposed method is beneficial over the other benchmarks. The estimated prediction error obtained for $q$ has a radius of $q_{0.95}=\QBound$ dB for 95\%. We tested this on separate no-mitigation evaluation points and observed that this interval covers \EvalCoverage{}\% of them, which is close to the 95\% target. The evaluation in Table~\ref{tab:main_results} shows that the proposed method achieves \PropThroughput{}~Mbit/s throughput and uses \PropEnergy{}~J across all \acp{RU}. The decision time \ac{RSRP} violation rate observed during the simulation is \PropViolation{}\%. Our proposed method tolerates \PropTolerate{}\%, defers \PropDefer{}\%, and mitigates \PropMitigate{}\% of the conflict batches. 

\begin{table}[t]
\centering
\caption{Closed-loop results over \EvalRuns{} held-out seeds.}
\label{tab:main_results}
\setlength{\tabcolsep}{3.2pt}
\begin{tabular}{lccc}
\toprule
Policy & Throughput (Mbit/s) & RU energy (J) & RSRP viol. (\%)\\
\midrule
No mitigation & \ObsThroughput & \ObsEnergy & \ObsViolation\\
Fixed priority & \PriThroughput & \PriEnergy & \PriViolation\\
Proposed & \PropThroughputCI & \PropEnergyCI & \PropViolationCI\\
\bottomrule
\end{tabular}
\end{table}

Afterwards, we compare each seed for the three different mitigation policies to obtain a clear picture rather than just an average. Table~\ref{tab:paired_delta} shows the difference between the proposed and reference policies for each seed, and then averages these results. We evaluated this way to remove the variation between seeds, irrespective of mobility randomness, and isolate the effect of the mitigation policy itself. Higher throughput and lower energy or violation values indicate better performance.

\begin{table}[t]
\centering
\caption{Paired KPI difference: proposed minus reference.}
\label{tab:paired_delta}
\setlength{\tabcolsep}{3.1pt}
\begin{tabular}{lccc}
\toprule
Reference & Throughput (Mbit/s) & RU energy (J) & Viol. (pp)\\
\midrule
No mitigation & \PropObsThroughputDelta & \PropObsEnergyDelta & \PropObsViolationDelta\\
Fixed priority & \PropPriThroughputDelta & \PropPriEnergyDelta & \PropPriViolationDelta\\
\bottomrule
\end{tabular}
\end{table}

The proposed method performs better than the benchmarks because of the mechanism to manage each intervention. Table~\ref{tab:interventions} shows the mitigation and temporary deferral outcomes for the proposed method separately for the same seed. For the priority-based approach, the \ac{CCO} has higher priority compared to \ac{ES}. Therefore, we observed that the priority method always overrides the control decision coming from \ac{ES}. We observed that the proposed method mitigates only \PropMitigateCount{} of the 400 (30\% fewer full interventions), defers \PropDeferCount{} for a control cycle, and tolerates \PropTolerateCount{}. When there is a deferred decision, a delay for observing in the next cycle is needed, and it is not executed on the radio yet. Therefore, we do not consider deferrals as interventions to the network. But if we do, the total drops only from \PriorityInterventionCount{} to \PropChangedCount{}, which is only a 4.75\% decrease. The \ac{KPI} gains above show that the proposed method is more efficient than the benchmarks. 

\begin{table}[t]
\centering
\caption{Conflict-decision outcomes on a selected seed.}
\label{tab:interventions}
\setlength{\tabcolsep}{2.1pt}
\begin{tabular}{lcccc}
\toprule
Policy & Unchanged & Deferred & Mitigated & Changed\\
\midrule
No mitigation & 400 (100\%) & 0 & 0 & 0\\
Fixed priority & 0 & 0 & 400 (100\%) & 400 (100\%)\\
Proposed & 19 (4.75\%) & 101 (25.25\%) & 280 (70\%) & 381 (95.25\%)\\
\bottomrule
\end{tabular}
\end{table}

The proposed method comparing to the no-mitigation policy, the paired differences are \PropObsThroughputDelta{}~Mbit/s in throughput, \PropObsEnergyDelta{}~J in energy, and \PropObsViolationDelta{} percentage points in violations. All of these favour the proposed method. This shows the importance of conflict mitigation in \ac{OpenRAN}, as network performance is improved when mitigation is done. Against fixed priority, the differences are \PropPriThroughputDelta{}~Mbit/s, \PropPriEnergyDelta{}~J, and \PropPriViolationDelta{} percentage points. Therefore, the proposed method is clearly better than fixed priority as well and improves all the \acp{KPI}. All these reported results are not from replaying the stored best scenarios. We used random scenarios with control decisions coming from various \acp{xApp} changing the radio state in the simulation. This then impacts the control decision coming in the next control cycle. We capture real feedback, mobility, and energy effects throughout the simulation to produce the final outcome. Both the no-mitigation and fixed-priority baselines use the same control-decision path as the proposed method.

\subsection{Ablation and negative cases}
We perform an ablation study to check how sensitive these results are to the \ac{KPI} prediction interval width. The predictor used in this study is a simple \ac{RSRP} predictor just to test the proposed pipeline. Figure~\ref{fig:q_calibration} shows how each coverage target maps to an error radius measured from the same \CalibSamples{} calibration residuals. We change this radius, $q$, from $q_{0.80}$ to $q_{0.99}$ on seeds 11 to 20 and rerun the full closed loop at each value. We report all the results obtained from these changes in Figure~\ref{fig:q_ablation}.

This ablation study separates two claims, such as the safety claim and the robustness claim. However, only the safety claim depends on the prediction accuracy. Let's assume that a tolerated batch is unsafe with probability at most $\alpha_k$. The safety claim here is that the probability doesn’t exceed $\alpha_k$ regardless of prediction accuracy. This is because $q$ in Eq.~\eqref{eq:conformal} is measured from the predictor's actual errors, and is not merely assumed. A less accurate predictor just gets a wider interval, not false confidence. However, decisiveness of the proposed method changes based on the interval width. Figure~\ref{fig:q_ablation}(a) shows that as coverage increases from 80\% to 99\% and $q$ grows from 2.31 to 3.46~dB, the tolerate rate drops and the defer rate rises. A more extreme test confirms that reclassifying the same number of reports with wider radii shows the decisive share drops from 88.5\% at $q=\QBound$~dB to 49.5\% at $q=10$~dB and 14\% at $q=12$~dB.  This confirms that a less accurate predictor not only shifts the tolerate or mitigate boundary. Figure~\ref{fig:q_ablation}(b) to (d) shows that the performance degradation is gradual when $q$ changes. The number of \ac{KPI} violations, throughput, and RU energy all change by only a few percent as $q$ nearly doubles. This shows that the proposed method is robust to prediction errors in real deployments.

However, this robustness depends on the exchangeability assumption discussed in Section~III. We acknowledge that a real deployment with lower \ac{KPI} prediction accuracy than we have experimented with in this study so far may degrade the decisiveness and the coverage guarantee. A wider $q$ only protects against errors of a size already seen in our current experimental setup. We also report where the method may fail, since a statistical guarantee is only useful if its exceptions are clear. An unsafe tolerate happens when a tolerated batch is followed by an RSRP value below $T_R$ in the same run; we observe \UnsafeTolerate{} such events. In \InfeasibleMitigations{} mitigation events, even removing every negative control decision cannot meet the RSRP constraint. Paired tests against no mitigation show throughput reductions in \ThroughputRegressionSeeds{} seeds, more RSRP violations in \ViolationRegressionSeeds{} seeds, and higher energy use in \EnergyIncreaseSeeds{} seeds out of \EvalRuns{}.

An important thing to note is that our current predictor is simple and doesn’t consider the shadowing variations and changes in serving-cell association for \ac{RSRP} prediction. These simplifications make RSRP predictable, so the simple predictor remains accurate and $q$ stays small. However, it may vary for denser and less predictable scenarios with complex urban deployment. Our implementation is only limited to a direct conflict analysis between \ac{CCO} and \ac{ES} conflicts in this study. We plan to extend this study further for more conflict types and diverse deployments in our future work. 



\section{Conclusion}
This paper proposes a reactive conflict mitigation method as a decision layer between existing \ac{CDC} and \ac{CMC}. The proposed method triages each conflicting control decision as tolerate, defer, or mitigate. We evaluate a direct conflict pair between \ac{CCO} and \ac{ES} over \ac{TXP} in the ns3-oran control loop. The evaluation shows that the proposed method takes 30\% fewer mitigation decisions than the fixed-priority benchmark while improving the involved \acp{KPI}. We perform an ablation study highlighting the importance of the prediction errors in the proposed triage and how the decision changes when the prediction error interval $q$ widens or shrinks. In the future, we plan to extend this work for diverse conflict types with larger deployed \ac{xApp} and sophisticated \ac{KPI} prediction models. 

\bibliographystyle{IEEEtran}
\bibliography{references}
\end{document}